\documentclass[a4paper,UKenglish,cleveref,autoref,thm-restate]{lipics-v2021}

\usepackage{xcolor}

\providecommand{\category}[1]{}

\title{ZOR filters: fast and smaller than fuse filters}
\titlerunning{ZOR filters}

\author{Antoine Limasset}{Univ. Lille, CNRS, Centrale Lille, UMR 9189 CRIStAL, F-59000 Lille, France}{}{https://orcid.org/0000-0002-0669-4141}{}

\authorrunning{Antoine Limasset}

\Copyright{Anonymous}



\category{}

\relatedversion{}

\supplement{\url{https://github.com/Malfoy/ZOR/}}

\acknowledgements{}

\nolinenumbers
\begin{document}

\maketitle

\begin{abstract}
Probabilistic membership filters support fast approximate membership queries with a controlled false-positive probability $\varepsilon$ and are widely used across storage, analytics, networking, and bioinformatics \cite{chang2008bigtable,dayan2018optimalbloom,broder2004network,harris2020improved,marchet2023scalable,chikhi2025logan,hernandez2025reindeer2}.
In the static setting, state-of-the-art designs such as XOR and  fuse filters achieve low overhead and very fast queries, but their peeling-based construction succeeds only with high probability, which complicates deterministic builds \cite{graf2020xor,graf2022binary,ulrich2023taxor}.

We introduce \emph{ZOR filters}, a deterministic continuation of XOR/fuse filters that guarantees construction termination while preserving the same XOR-based query mechanism.
ZOR replaces restart-on-failure with deterministic peeling that abandons a small fraction of keys, and restores false-positive-only semantics by storing the remainder in a compact auxiliary structure.
In our experiments, the abandoned fraction drops below $1\%$ for moderate arity (e.g., $N\ge 5$), so the auxiliary handles a negligible fraction of keys.
As a result, ZOR filters can achieve overhead within $1\%$ of the information-theoretic lower bound $\log_2(1/\varepsilon)$ while retaining fuse-like query performance; the additional cost is concentrated on negative queries due to the auxiliary check.
Our current prototype builds several-fold slower than highly optimized fuse builders because it maintains explicit incidence information during deterministic peeling; closing this optimisation gap is an engineering target.
\end{abstract}

\section{Introduction}
\label{sec:Introduction}

Probabilistic membership filters are widely deployed data structures for representing sets and supporting approximate membership queries.
They appear in database and storage engines to avoid disk reads and unnecessary probes \cite{chang2008bigtable,o1996log,lakshman2010cassandra,dayan2018optimalbloom}, in distributed query processing and cloud analytics to reduce communication and remote lookups \cite{lohman1988r,ramesh2008optimizing,lee2012join,groppe2014p}, and in networked systems and caching protocols to summarize large sets under tight bandwidth constraints and support packet-processing tasks \cite{broder2004network,fan2000summary,snoeren2001hash,dharmapurikar2003longest,dharmapurikar2006fast}.
They also arise in bandwidth-sensitive blockchain light-client protocols \cite{qin2019applying}, and are pervasive in bioinformatics pipelines and indexes, where membership queries over $k$-mers and derived keys are a core primitive \cite{harris2020improved,marchet2023scalable,chikhi2025logan,hernandez2025reindeer2}.
Allowing a controlled false-positive probability $\varepsilon$ yields major space savings relative to exact representations, which must store full keys and typically incur load-factor overhead.
A standard reference point is the information-theoretic bound of $\log_2(1/\varepsilon)$ bits per stored key.
Practical designs are therefore often compared by their multiplicative or additive overhead above this bound.

Bloom filters remain a strong baseline \cite{bloom1970space,tarkoma2011theory} for such usage.
Their appeal comes from a simple interface, linear-time construction, constant-time queries, and straightforward dynamic usage.
With optimal tuning, a Bloom filter storing $m$ bits for $n$ keys and using $k$ hash functions has false-positive rate
\[
\varepsilon \approx \left(1-e^{-kn/m}\right)^k
\]
With the optimal number of hash functions 
\[
k^\star \approx \frac{m}{n}\ln 2,
\]
this implies the well-known space bound
\[
\frac{m}{n} \approx \frac{-\ln \varepsilon}{(\ln 2)^2} \approx 1.44\,\log_2(1/\varepsilon)\ \ \text{bits per key}.
\]
Thus, Bloom filters incur about $44\%$ overhead in bits per key relative to $\log_2(1/\varepsilon)$ \cite{tarkoma2011theory}.
In practice, Bloom filters  face a tension between accuracy and query throughput.
Achieving smaller $\varepsilon$ requires increasing $m/n$ and, at the optimum, increasing $k^\star \approx (m/n)\ln 2$, which increases the number of probed locations.
On modern CPUs, random memory access and cache misses dominate latency, so these probes can become the bottleneck.
A common mitigation is the blocked Bloom filter family, which restricts all probed positions to a single cache-line-sized block to improve locality and throughput \cite{putze2007cache,lang2019performance}.
Blocking reduces entropy and introduces uneven occupancy across blocks, increasing the false-positive rate at fixed $m/n$.
Equivalently, maintaining the same false-positive probability requires allocating more bits per key.
For example, for a target false-positive rate around $1\%$, Lang \emph{et al.} report that a classic Bloom filter requires roughly $10$ bits per key, whereas aggressive register-blocked designs require roughly $12$--$14$ bits per key, corresponding to about $20$--$40\%$ additional space for the same error rate \cite{lang2019performance}.

Alternative dynamic filters have therefore attracted substantial attention, with cuckoo filters being a prominent example \cite{fan2014cuckoo}.
By storing short fingerprints in a cuckoo-hashed table, cuckoo filters typically answer queries with as few as two cache-line accesses while supporting deletions.
Their space efficiency depends on occupancy, bucket size, and fingerprint length, and is commonly tens of percent above $\log_2(1/\varepsilon)$ once metadata and load-factor slack are accounted for \cite{fan2014cuckoo,eppstein2016cuckoo}.
Beyond cuckoo-style relocation, quotient-based filters (quotient filters, Morton filters, and vector quotient filters) emphasize contiguous or structured layouts that limit random probing by storing short remainders together with compact metadata \cite{bender2011don,breslow2018morton,pandey2021vector}.
Compared to cuckoo filters, which probe a small number of candidate buckets and may relocate items on insertion, quotient-based designs typically resolve membership by scanning short runs determined by the quotient and by exploiting cache- and SIMD-friendly encodings.
Overall practical dynamic designs typically pay for metadata and load-factor slack, leaving them noticeably above the $\log_2(1/\varepsilon)$ bound in many regimes.

By contrast, static key sets, built once and then queried repeatedly without updates, are common in practice, notably in immutable storage components such as SSTables in LSM-tree engines, read-only reference dictionaries and indexing side structures, and large bioinformatics $k$-mer indexes \cite{o1996log,marchet2021blight,pibiri2023weighted,pibiri2023locality}.
In this regime, substantially tighter space is achievable.
Many static approximate membership filters are closely connected to retrieval structures and minimal perfect hash functions (MPHFs) \cite{lehmann2025modern}.
A straightforward construction is to build an MPHF $h$ that maps each key to a unique index, store a short fingerprint at that index, and answer membership queries by recomputing the index and comparing fingerprints \cite{dietzfelbinger2008succinct,marchet2020resource}.
In this paradigm, the main overhead is the MPHF itself.
While the information-theoretic minimum is $\log_2(e)\approx 1.44$ bits per key, practical builders still contribute a small additive constant \cite{pibiri2021pthash,limasset2017bbhash,hermann2024phobic,lehmann2025modern}.
Because this cost is additive and independent of $\varepsilon$, it is negligible when fingerprints are long but becomes a large relative overhead when fingerprints are short.
For instance, a $2$-bit MPHF overhead adds $25\%$ space with $8$-bit fingerprints, $12.5\%$ with $16$-bit fingerprints, and $6.25\%$ with $32$-bit fingerprints.

Recent static filters such as XOR filters, Ribbon filters, and fuse filters push this principle further. 
Inspired by MPHF-like placement ideas, they offer linear-time construction and very fast queries, while achieving low multiplicative overheads above $\log_2(1/\varepsilon)$ \cite{graf2020xor,dillinger2021ribbon,graf2022binary}.
In particular, state-of-the-art fuse filters achieve theoretical storage of $1.125\log_2(1/\varepsilon)$ bits per key in the 3-wise case and $1.075\log_2(1/\varepsilon)$ in the 4-wise case, corresponding to about $12.5\%$ and $7.5\%$ overhead, respectively \cite{graf2022binary}.
These filters are attractive because they combine near-optimal space, high query throughput with few memory accesses (typically less than $100\,\mathrm{ns}$), and fast construction suitable for large-scale pipelines.

In this manuscript, we propose \emph{ZOR filters}, a continuation of XOR/fuse filters that guarantees construction termination while preserving the same XOR-based query mechanism.
ZOR eliminates restart-on-failure and reduces space overhead toward the information-theoretic limit by abandoning a small remainder set and handling it with a compact auxiliary structure, thereby restoring false-positive-only semantics.
We analyse the resulting space/error trade-off and empirically evaluate abandonment rates, memory overhead, and query/build performance across arities and segment sizes.

\section{Methods}
\label{sec:Methods}

\subsection{Preliminaries: fuse-filter construction}
We briefly recap fuse filter construction to fix notation and to separate prior work \cite{graf2022binary} from the contributions of this manuscript.

We consider a static set $S$ of $n$ distinct keys and an array of $m$ cells. 
Each key $x\in S$ is mapped, via hashing, to $N$ cells
\[
H(x)=\{h_1(x),h_2(x),\dots,h_N(x)\}\subseteq \{0,\dots,m-1\}.
\]
The filter stores in each cell an $F$-bit value (a ``cell fingerprint'') and answers a membership query for $y$ by combining the $N$ cell values at $H(y)$ by bitwise XOR and comparing against a fingerprint of $y$ \cite{graf2020xor,graf2022binary}.
Fuse filters differ from XOR filters primarily in how $H(x)$ is chosen to improve locality and construction speed. The array is partitioned into many disjoint segments, and $H(x)$ consists of one location in each of $N$ consecutive segments. In practice segments are sized as powers of two to accelerate range reduction. This ``consecutive-segment'' constraint improves cache behavior during both query and construction and also reduces the required overhead ($m/n$) for successful peeling.

Construction proceeds in two stages. First, each cell records which keys map to it: for every $x\in S$, we insert $x$ into the incidence lists of all cells in $H(x)$. Equivalently, this defines an  hypergraph whose vertices are the $m$ cells and whose hyperedges are the keys, each hyperedge connecting the $N$ vertices $H(x)$. 
Second, the algorithm performs a peeling process. We iteratively search for any cell $v$ that currently contains a unique key (i.e., whose current degree is $1$). If such a cell exists, let $x$ be its unique incident key. We link $x$ to $v$ (recording that $x$ will later be resolved at $v$), and remove all occurrences of $x$ from the other cells in $H(x)\setminus\{v\}$, updating their degrees. This removal can create new cells of degree $1$, which are then processed in the same way. The procedure continues until either every key has been linked, or no cell of degree $1$ exists while unprocessed keys remain; in that case the construction fails and is typically restarted with different hash seeds.

Once a full linking order exists (i.e., all keys have been peeled), the cell payloads are assigned in reverse order. Let $f(\cdot)$ be an $F$-bit fingerprint function. When processing a key $x$ linked to cell $v\in H(x)$, all other incident cell values have already been fixed; we set
\[
T[v] \leftarrow f(x)\ \oplus\ \bigoplus_{u\in H(x)\setminus\{v\}} T[u],
\]
so that $\bigoplus_{u\in H(x)}T[u]=f(x)$. This completes the construction.

\subsection{ZOR construction: deterministic peeling by abandoning keys}
The core issue of XOR/fuse peeling is that the process can get stuck on a remaining subgraph in which every still-active cell has degree at least $2$ (i.e., no degree-$1$ cell exists), which forces a complete restart in the standard algorithm.
Our solution avoids restarts by using an always-terminating variant that \emph{abandons} keys whenever the process blocks.
Allowing a controlled false-negative rate has precedent in other filtering settings, e.g., stable Bloom filters for streaming where older items are intentionally forgotten \cite{deng2006approximately} and retouched Bloom filters that trade false negatives for fewer false positives \cite{donnet2006retouched}.
Our main goal here is different: we use such filters as an intermediate object and then recover standard false-positive-only semantics via an auxiliary structure.

Formally, we follow the fuse construction algorithm, but we modify the step taken when the queue of degree-$1$ cells becomes empty while keys remain. 
Instead of restarting, we select a cell $v$ with minimal current degree
\[
d_{\min}=\min_{u} d(u), \qquad d_{\min}\ge 2,
\]
and consider the $d_{\min}$ keys currently incident to $v$. 
We keep one key $x^\star$ (to be resolved at $v$) and abandon the other $d_{\min}-1$ keys by removing them from all their incident cells. 
After this forced removal, cell $v$ becomes degree $1$ and the usual peeling step can proceed without restarting.
This forced removal decreases degrees and guarantees progress. 
Repeating this rule ensures termination because each step removes at least one still-active key from the incidence structure.

Let $A\subseteq S$ denote the abandoned set and $\alpha=|A|/n$ the abandoned fraction. 
The main structure is then constructed exactly as a fuse filter, but only for $S\setminus A$. 
This main structure alone is an approximate filter with both false positives (as fuse filter) and false negatives (on keys in $A$). 
We refer to this single-stage object (without an auxiliary) as a \emph{pure ZOR filter}.
Pure ZOR filters are interesting in their own right when a very small false-negative rate is acceptable (in our measurements often below the $1\%$ scale), but the main goal of this work is to recover the standard false-positive-only semantics.
We propose to achieve this by explicitly handling $A$ with an auxiliary structure.

A practical consequence of always-terminating construction is that it removes restart-driven uncertainty: for a fixed array size $m$, construction never fails, and the ``difficulty'' of the instance manifests as $\alpha$ rather than as a rebuild probability. 
In particular, if one chooses $m=n$ (one cell per key), then the main structure uses exactly $F$ bits per original key; choosing $m>n$ reduces the abandoned fraction at the cost of a multiplicative space factor $m/n$, whereas choosing $m<n$ yields a cheaper main structure but necessarily increases false negatives.
In the following, we choose $m=n$ because it minimises the main-structure footprint for our intended use; we leave the study of other $m/n$ trade-offs to future work.

\subsection{Reducing the abandoned fraction: choosing the intervention cell}
When the peel queue is empty, the algorithm must force progress on a remaining core in which all active cells have degree at least $2$.
It does so by selecting an \emph{intervention cell} $v$ and processing all currently active keys incident to $v$: exactly one key is kept and will be resolved at $v$, while the remaining $d(v)-1$ incident keys are abandoned and removed from the incidence structure.
A consequence is that, once $v$ is chosen, \emph{all} keys incident to $v$ disappear from the active structure (either as kept or abandoned).
Therefore, the subsequent evolution of degrees is governed mainly by the choice of the intervention cell $v$, the choice of which single key to keep affects only which key remains in $S\setminus A$ (and hence which key does not require auxiliary handling).

Accordingly, we choose $v$ among currently minimal-degree cells so as to minimise the number of abandoned keys created by each intervention.
This is a local choice based only on the current degrees and incidences, and it does not guarantee a globally minimal abandoned set.
Let $\mathcal{I}(v)$ be the set of currently active keys incident to $v$, and let $d(u)$ be the current degree of a cell $u$.
Intervening at $v$ removes every key in $\mathcal{I}(v)$ from each of its other incident cells, i.e., from cells in
\[
\mathcal{N}(v)=\{u\neq v : \exists x\in \mathcal{I}(v)\ \text{s.t.}\ u\in H(x)\}.
\]
A simple proxy for how much this intervention will ``unlock'' the peel process is how many degree-$2$ neighbours are hit, since decreasing a degree-$2$ cell by one  creates a new degree-$1$ cell.
We capture this with
\[
\mathrm{deg2}(v)=\sum_{x\in \mathcal{I}(v)} \bigl|\{u\in H(x)\setminus\{v\}: d(u)=2\}\bigr|.
\]

When multiple candidates have similar $\mathrm{deg2}(v)$, we use secondary scores to prefer interventions whose impact is concentrated in lighter neighbourhoods.
With multiplicity over incidences, define
\[
\mathrm{sum}(v)=\sum_{x\in \mathcal{I}(v)}\ \sum_{u\in H(x)\setminus\{v\}} d(u),
\qquad
\mathrm{max}(v)=\max_{x\in \mathcal{I}(v)}\ \max_{u\in H(x)\setminus\{v\}} d(u),
\]
and let $\mathrm{vec}(v)$ be the multiset of degrees $\{d(u): u\in H(x)\setminus\{v\},\ x\in \mathcal{I}(v)\}$ sorted in non-decreasing order.

In our implementation, when a blocking event occurs we scan up to $T$ distinct minimal-degree cells.
For each scanned candidate $v$ we compute the scores above, then pick the best-scoring candidate under a chosen policy.
This parameter $T$ controls the extra work spent per blocking event.
We implemented the following intervention policies (all restricted to cells of minimal current degree, or to the best among the $T$ scanned minimal-degree candidates):
\begin{itemize}
    \item \emph{Lightest-neighbourhood:} minimise $\mathrm{sum}(v)$; break ties by larger $\mathrm{deg2}(v)$, then smaller $\mathrm{max}(v)$.
    \item \emph{Heaviest-neighbourhood:} maximise $\mathrm{sum}(v)$; break ties by smaller $\mathrm{deg2}(v)$, then larger $\mathrm{max}(v)$.
    \item \emph{MostDeg2:} compare $\mathrm{vec}(v)$ lexicographically (preferring smaller degrees earlier); break ties by larger $\mathrm{deg2}(v)$, then smaller $\mathrm{sum}(v)$ and smaller $\mathrm{max}(v)$.
    \item \emph{MinMaxDegree:} minimise $\mathrm{max}(v)$; break ties by smaller $\mathrm{sum}(v)$, then larger $\mathrm{deg2}(v)$.
\end{itemize}
Once $v$ is selected, the choice of the single kept key among $\mathcal{I}(v)$ is made deterministically (e.g., by a fixed hash order) for reproducibility.

\subsection{Handling abandoned keys}
Pure ZOR filters may return false negatives precisely on the abandoned set $A$.
To recover the standard false-positive-only semantics, we store the abandoned keys that would otherwise be missed by the main structure in an auxiliary static structure, such as a fuse filter or an MPHF-plus-fingerprint table.
Concretely, we define $A'\subseteq A$ as the subset of abandoned keys for which the main structure returns ``absent'' at build time (see Section~\ref{sec:opportunistic-skip} for more details); storing $A'$ suffices to eliminate false negatives.

Let the main filter have false-positive probability $\varepsilon_1$ and the auxiliary have false-positive probability $\varepsilon_2$.
We answer membership queries by checking the main structure first and consulting the auxiliary only if needed.
A query returns ``present'' if either structure matches, hence the overall false-positive probability is
\[
\begin{aligned}
\varepsilon_{\mathrm{tot}}
&= 1-(1-\varepsilon_1)(1-\varepsilon_2) \\
&= \varepsilon_1+\varepsilon_2-\varepsilon_1\varepsilon_2 \\
&\le \varepsilon_1+\varepsilon_2,
\end{aligned}
\]
and the difference is $\varepsilon_1\varepsilon_2$, which is negligible when $\varepsilon_1,\varepsilon_2\ll 1$.
The next subsection selects the auxiliary fingerprint size to optimise memory efficiency.

\subsection{Memory-optimal auxiliary fingerprint size}
A natural question is how large the auxiliary fingerprint should be to minimise space at a target overall error.
For this analysis, we ignore implementation-specific constants and measure only fingerprint bits.
The main filter stores $F$ bits per original key when $m=n$, the auxiliary stores $G$ bits per auxiliary-stored key, and the abandoned fraction is $\alpha$, so the total bits per original key are
\[
B(F,G)=F+\alpha G.
\]
Under the standard fingerprint model, $\varepsilon_1\simeq 2^{-F}$ and $\varepsilon_2\simeq 2^{-G}$, and therefore
\[
\varepsilon_{\mathrm{tot}}(F,G)=1-(1-2^{-F})(1-2^{-G})
\approx 2^{-F}+2^{-G}.
\]
Define the multiplicative overhead relative to the information-theoretic bound as
\[
\rho(F,G)=\frac{B(F,G)}{-\log_2\!\bigl(\varepsilon_{\mathrm{tot}}(F,G)\bigr)}.
\]
Optimizing under the approximation $\varepsilon_{\mathrm{tot}}\approx 2^{-F}+2^{-G}$ yields the balance condition
\[
2^{-G^\star}\approx \alpha\,2^{-F}
\qquad\Longleftrightarrow\qquad
G^\star \approx F+\log_2\!\Bigl(\frac{1}{\alpha}\Bigr),
\]
with a detailed derivation provided in Appendix~\ref{app:aux-opt}.

Thus, the auxiliary false-positive rate should be on the order of $\alpha$ times the main false-positive rate.
In practice we take the ceiling, for example
\[
G^\star \approx F+\Bigl\lceil \log_2\!\Bigl(\frac{1}{\alpha}\Bigr)\Bigr\rceil.
\]
For $\alpha=1\%$, $\log_2(1/\alpha)=\log_2(100)\approx 6.64$, suggesting $G^\star\approx F+7$.
In our experiments, $\alpha$ is often below $1\%$ for moderate arity (e.g., $N\ge 5$), so we use $G=F+8$ as a simple default that keeps $\varepsilon_2$ negligible, remaining close to optimal while being efficient in practice.

\subsection{Opportunistic skip}
\label{sec:opportunistic-skip}

Before inserting an abandoned key $x\in A$ into the auxiliary filter, we may query the main filter on $x$. 
If the main filter already returns ``present'' for $x$, then $x$ will not be a false negative at query time and does not require auxiliary handling. Under a uniform fingerprint model, this happens with probability approximately $\varepsilon_1$. 
This optimisation is negligible for low $\varepsilon_1$ (large fingerprints) but can be noticeable for small $F$ (especially below 8 bits).
Overall, this optimisation can reduce the auxiliary size, especially when $F$ is small, at the modest build-time cost of one extra main-filter query per abandoned key.

\subsection{Cascading auxiliary stages and practical limits}
Instead of using a single auxiliary filter, a natural idea would be to  cascade stages: build a primary ZOR structure on $S$, then build a second-stage structure on the abandoned keys, and repeat until the remainder is small enough to be handled by a different representation (e.g., a tiny exact set or a final conventional filter).
If the first stage abandons a fraction $\alpha_1$ of the original keys and the second stage abandons a fraction $\alpha_2$ of its input, then the size of the remaining set after two stages is $\alpha_1\alpha_2 n$.
Consequently, the space contribution of later stages shrinks multiplicatively: a third stage contributes a term proportional to $\alpha_1\alpha_2$, and so on.

Cascading can therefore reduce the auxiliary footprint, but the gain is typically limited when $\alpha_1$ is already small, while query time increases because negative queries may need to check multiple stages.
This trade-off is particularly clear in our current design, where the auxiliary is implemented as a 4-wise  fuse filter with multiplicative overhead $1.075$ over the information-theoretic bound \cite{graf2022binary}.
With a main ZOR fingerprint of $F$ bits and auxiliary fingerprints of $F+8$ bits, the auxiliary contributes an overhead (relative to $F$) of approximately
\[
\Delta_{\mathrm{ZOR+FUSE}} \approx   \alpha_1 \cdot 1.075 \cdot \frac{F+8}{F}\ \  .
\]
For instance, with $\alpha_1=0.5\%$ and $F=16$, this evaluates to
\[
\Delta_{\mathrm{ZOR+FUSE}} \approx   0.005 \cdot 1.075 \cdot \frac{24}{16} \approx 0.81\%.
\]

If one adds a second ZOR stage before the final fuse auxiliary (i.e., ZOR+ZOR+FUSE), and if the second stage can itself be implemented at about $1\%$ overhead over its own information-theoretic target, then the effective overhead factor for handling the abandoned keys becomes roughly $1.01$ instead of $1.075$ for that portion.
This yields the approximation
\[
\Delta_{\mathrm{ZOR+ZOR+FUSE}} \approx  \alpha_1 \cdot 1.01 \cdot \frac{F+8}{F}\ \  ,
\]
which for the same $\alpha_1=0.5\%$ and $F=16$ gives
\[
\Delta_{\mathrm{ZOR+ZOR+FUSE}} \approx  0.005 \cdot 1.01 \cdot \frac{24}{16} \approx 0.76\%.
\]
Thus, cascading replaces an overhead around $0.81\%$ by about $0.76\%$ in this representative regime: the absolute gain is small because the auxiliary is already weighted by $\alpha_1$.
This is the main practical limitation of cascading in our setting: once $\alpha_1$ is below the percent level, additional stages offer diminishing space returns  but impose a direct cost in query time, since negatives may need to evaluate more stages before rejecting.
Accordingly, we use a single auxiliary stage in the experiments and treat cascading mainly as an optional refinement for cases where query-time budgets are less stringent or when most queries are positive.
Moreover, even if the abandoned keys were handled at the information-theoretic limit (i.e., with exactly $G$ bits per abandoned key to match a $2^{-G}$ target), the auxiliary would still cost at least $\alpha G$ additional bits per original key, so the multiplicative space overhead has a hard lower bound of $\alpha$.
In other words, once $\alpha$ is already small, the dominant lever for further space improvements is not adding stages but reducing $\alpha$ itself (e.g., by increasing arity, tuning segment parameters, or improving the abandonment policy), because every auxiliary term is multiplied by $\alpha$ and therefore exhibits diminishing returns from additional cascading.

\subsection{Construction burden}

A key to the extremely fast fuse-filter construction is that the builder can avoid explicit adjacency lists during peeling because the algorithm only ever removes degree-$1$ cells.
Each cell can maintain just two aggregates: a degree counter and the XOR of the  hashes of incident keys.
When a cell reaches degree $1$, the XOR aggregate reveals the unique remaining key, enabling the builder to continue without knowing the full incidence list. 
If peeling gets stuck in a core, fuse construction simply restarts with a new seed, so it never needs to identify and delete a specific key from a multi-degree cell.

During ZOR construction when the peel queue is empty, it must choose a key to abandon from a cell with degree at least $2$. 
At that moment, the count+XOR trick is insufficient because it does not identify which active key should be removed, nor does it support enumerating the incident keys. 
The builder therefore needs per-cell membership information (an adjacency list or an equivalent structure) to locate an active key and update all its incident cells. 
This additional structure increases memory traffic and random access during peeling, which explains why the current ZOR implementation is slower to build even though query evaluation remains similar to fuse filters.

\section{Results}
\label{sec:Results}

We now evaluate ZOR filters along three axes: the abandoned fraction $\alpha$ induced by deterministic peeling,  end-to-end memory cost when abandoned keys are handled by an auxiliary structure, and  performance trade-offs in construction and querying.

\begin{figure}
    \centering
    \includegraphics[width=0.49\linewidth]{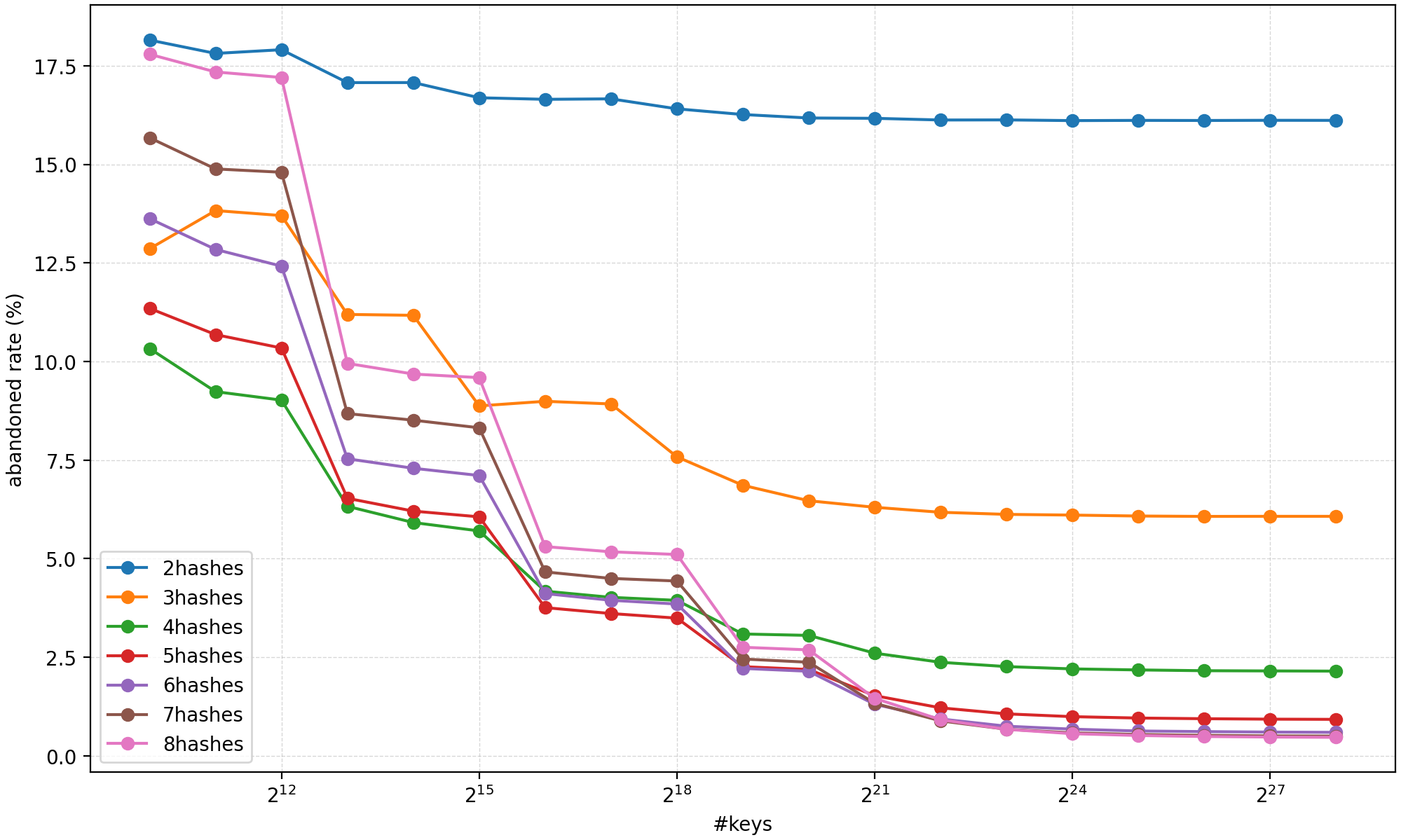}
    \includegraphics[width=0.49\linewidth]{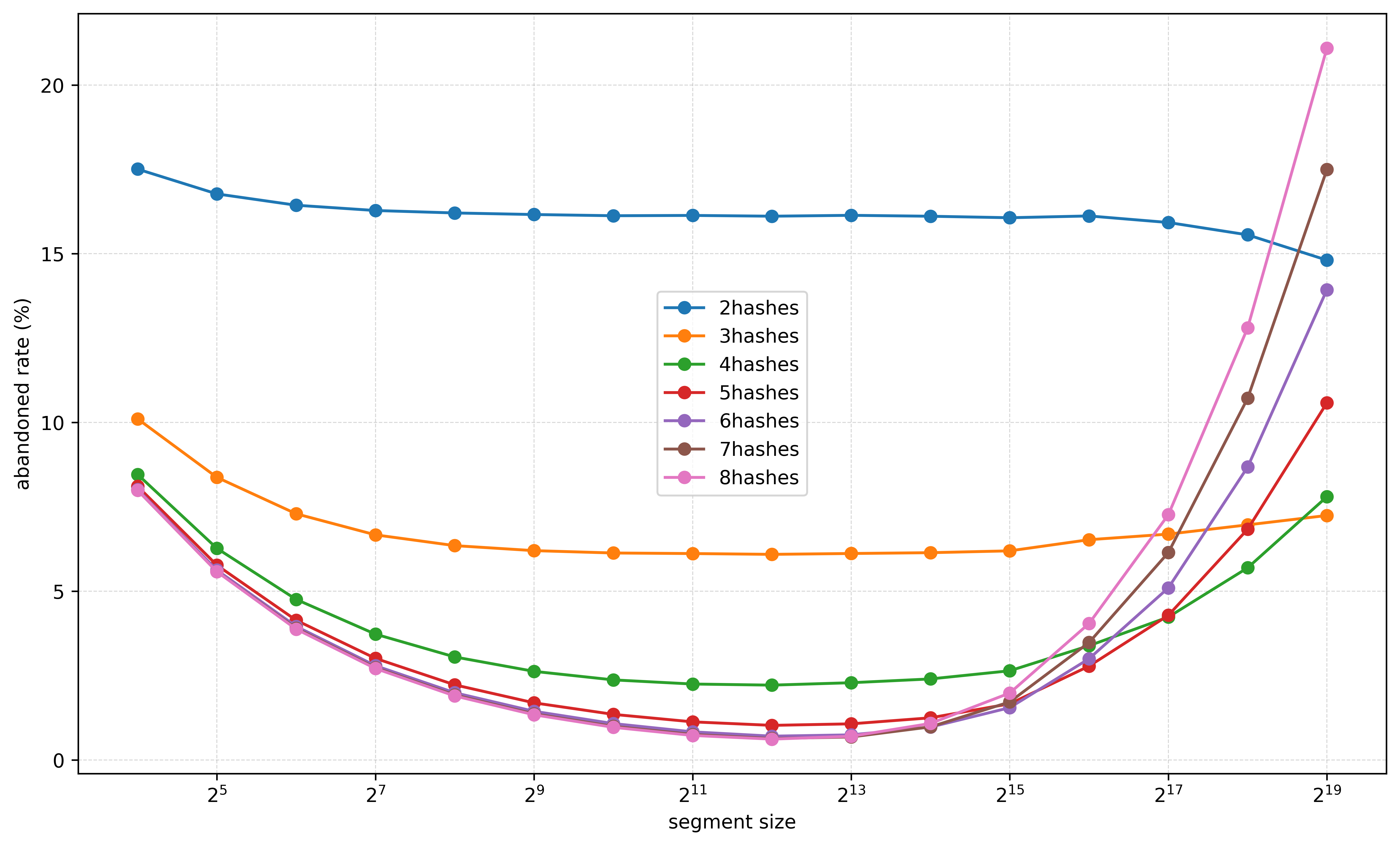}
    \caption{
    Abandoned fraction $\alpha$ as a function of arity $N$ for several set sizes $n$ (left).
    Abandoned fraction $\alpha$ as a function of the segment (block) size for several arities $N$ (right).
    }
    \label{fig:Abandoned}
\end{figure}

We start by measuring the abandoned fraction $\alpha$ as a function of the arity $N$ (number of hash functions) and the set size $n$ (Figure~\ref{fig:Abandoned}, left).
Increasing $N$ reduces $\alpha$ up to diminishing returns above $5$, and larger sets tend to improve behavior, consistent with prior observations for XOR- and fuse-style peeling constructions.
For the largest sets we tested, the measured abandonment rates were $6.1\%$ (3-wise), $2.1\%$ (4-wise), $0.9\%$ (5-wise), $0.6\%$ (6-wise), $0.5\%$ (7-wise), and $0.4\%$ (8-wise).
Thus, abandonment below $1\%$ is achievable in practice with $N\ge 5$, so the auxiliary handles at most $\alpha n$ keys, about two orders of magnitude fewer keys than the main structure.
In the pure ZOR setting (without an auxiliary), these values also correspond directly to very low false-negative rates.

We also study the segment (block) size used for fuse-style hashing (Figure~\ref{fig:Abandoned}, right).
As in fuse filters, we observe an intermediate segment size that minimises abandonment, reflecting a trade-off between locality (small segments) and effective randomness in the induced hypergraph (large segments).

\begin{figure}
    \centering
    \includegraphics[width=0.49\linewidth]{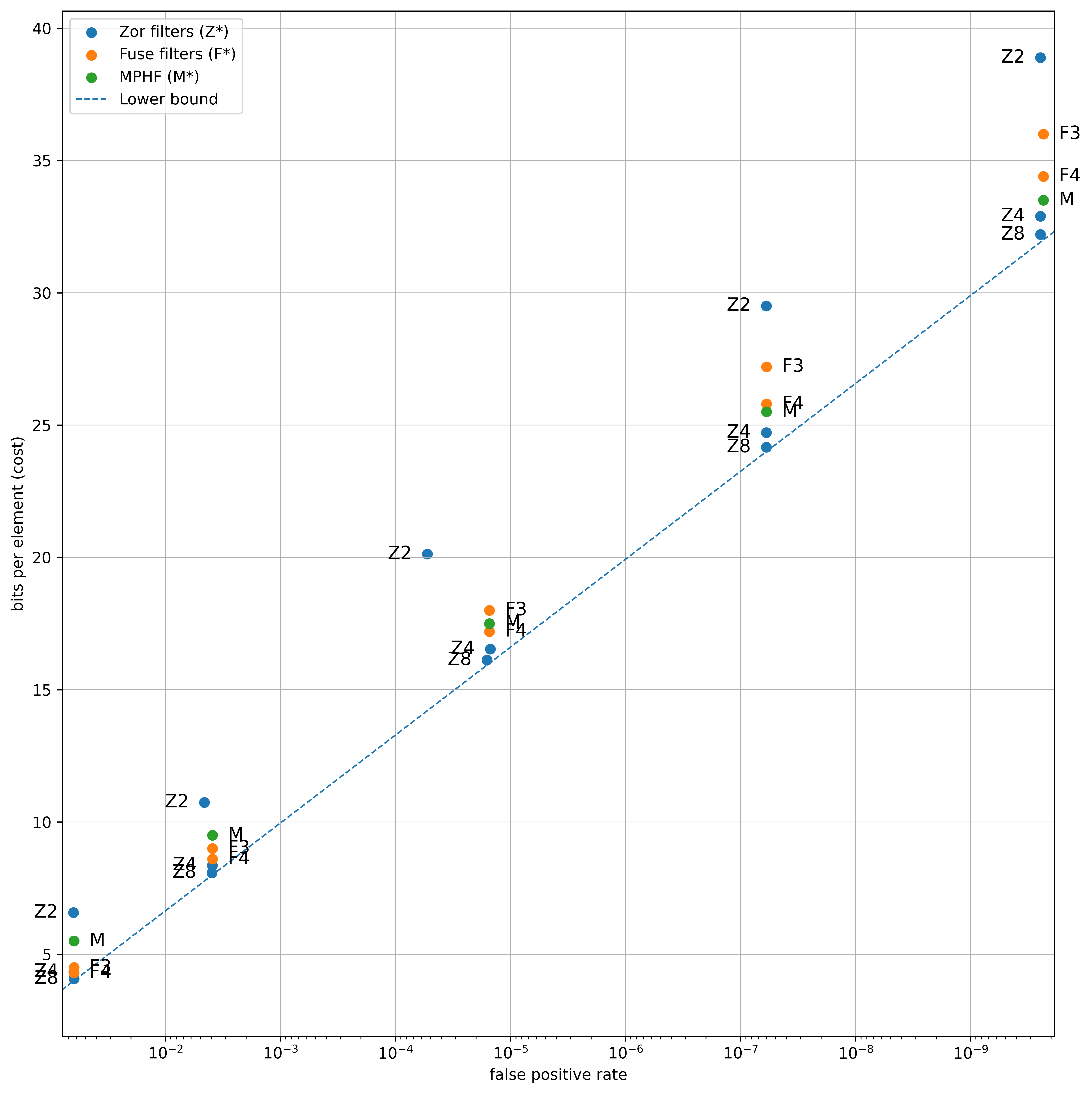}
    \includegraphics[width=0.49\linewidth]{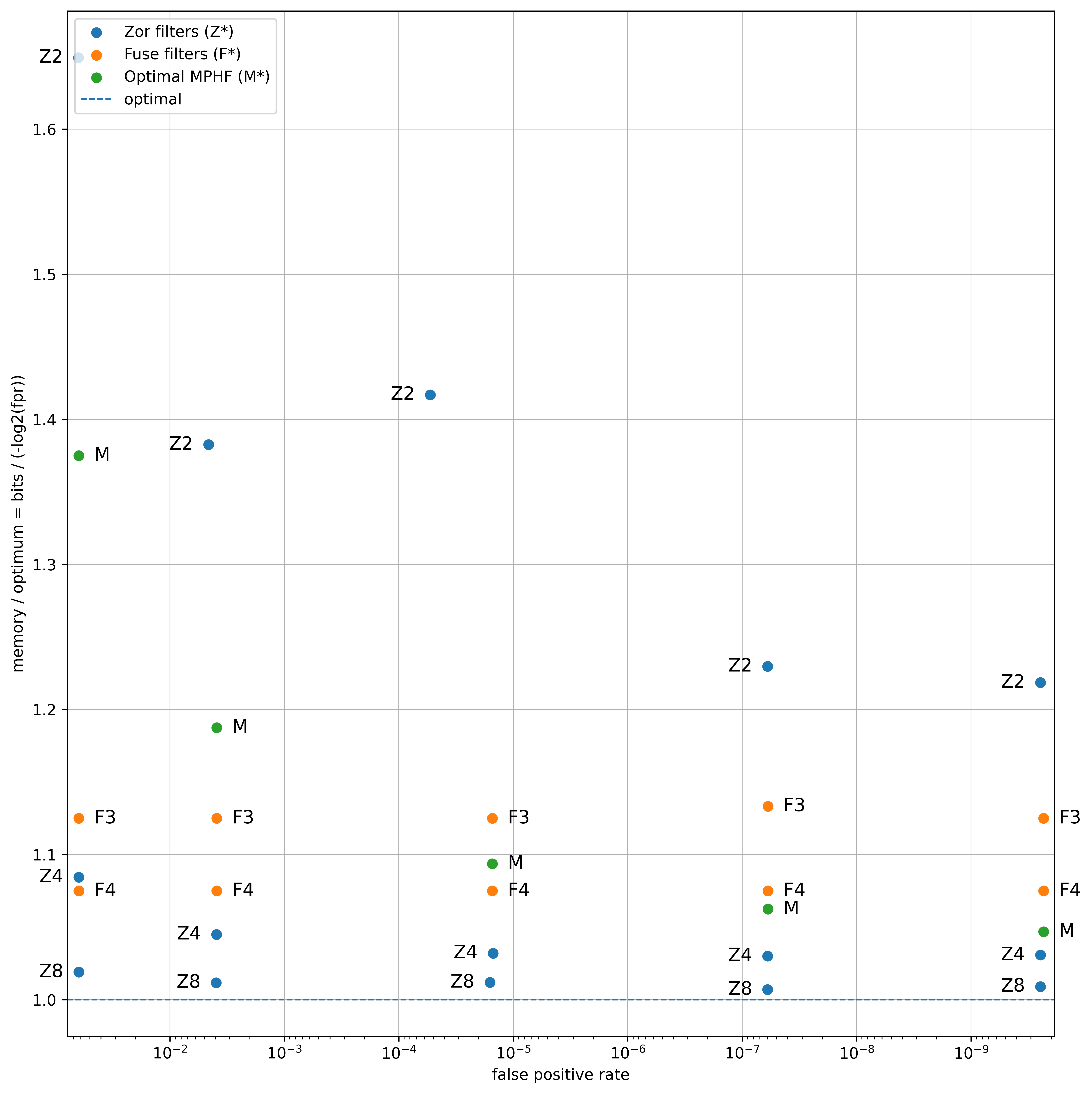}
\caption{Memory efficiency of  fuse filters, ZOR filters, and an MPHF+fingerprint baseline (MPHF cost fixed to the information-theoretic minimum $1.44$ bits/key). Fingerprint sizes of $4$, $8$, $16$, $24$, and $32$ bits are tested.
For ZOR, the reported false-positive rate accounts for both the main and auxiliary structures.}

    \label{fig:Sizes}
\end{figure}

We then compare the total memory cost of ZOR filters against state-of-the-art static baselines:  fuse filters and an information-theoretic MPHF+fingerprint baseline in which the MPHF is assumed to meet the lower bound, costing $\log_2(e)\approx 1.44$ bits per element (Figure~\ref{fig:Sizes}).
For simplicity, our implementation uses a 4-wise fuse filter as the auxiliary structure; depending on the fingerprint regime, an MPHF-based auxiliary could also be considered. We also report (Figure~\ref{fig:Sizes}, left) the memory cost per element for various fingerprint sizes and (Figure~\ref{fig:Sizes}, right) the corresponding overhead relative to $\log_2(1/\varepsilon_{\mathrm{tot}})$, where $\varepsilon_{\mathrm{tot}}$ accounts for the combined false-positive rate of the main and auxiliary structures.
Fuse filters are very competitive, with overhead of about $7.5\%$ in the 4-wise case \cite{graf2022binary}.
The MPHF baseline has an additive overhead (in practice often around the $1.44$--$2$ bits/key scale depending on the construction), so it becomes increasingly attractive for larger fingerprints \cite{lehmann2025modern,pibiri2021pthash}.
Across all tested fingerprint sizes, ZOR filters improve over single-stage fuse filters at the same main arity $N$, and higher-arity ZOR (e.g., 8-wise) further reduces overhead.
The relative benefit is strongest for larger fingerprints because the auxiliary uses $G=F+8$ bits and the auxiliary cost $\alpha G$ becomes (relatively) small as $F$ grows. 
In our measurements, ZOR surpasses the MPHF baseline even in the $32$-bit fingerprint regime, and we observe overheads below the $1\%$ scale, extremely close to the information-theoretic limit.

\begin{figure}
    \centering
    \includegraphics[width=0.49\linewidth]{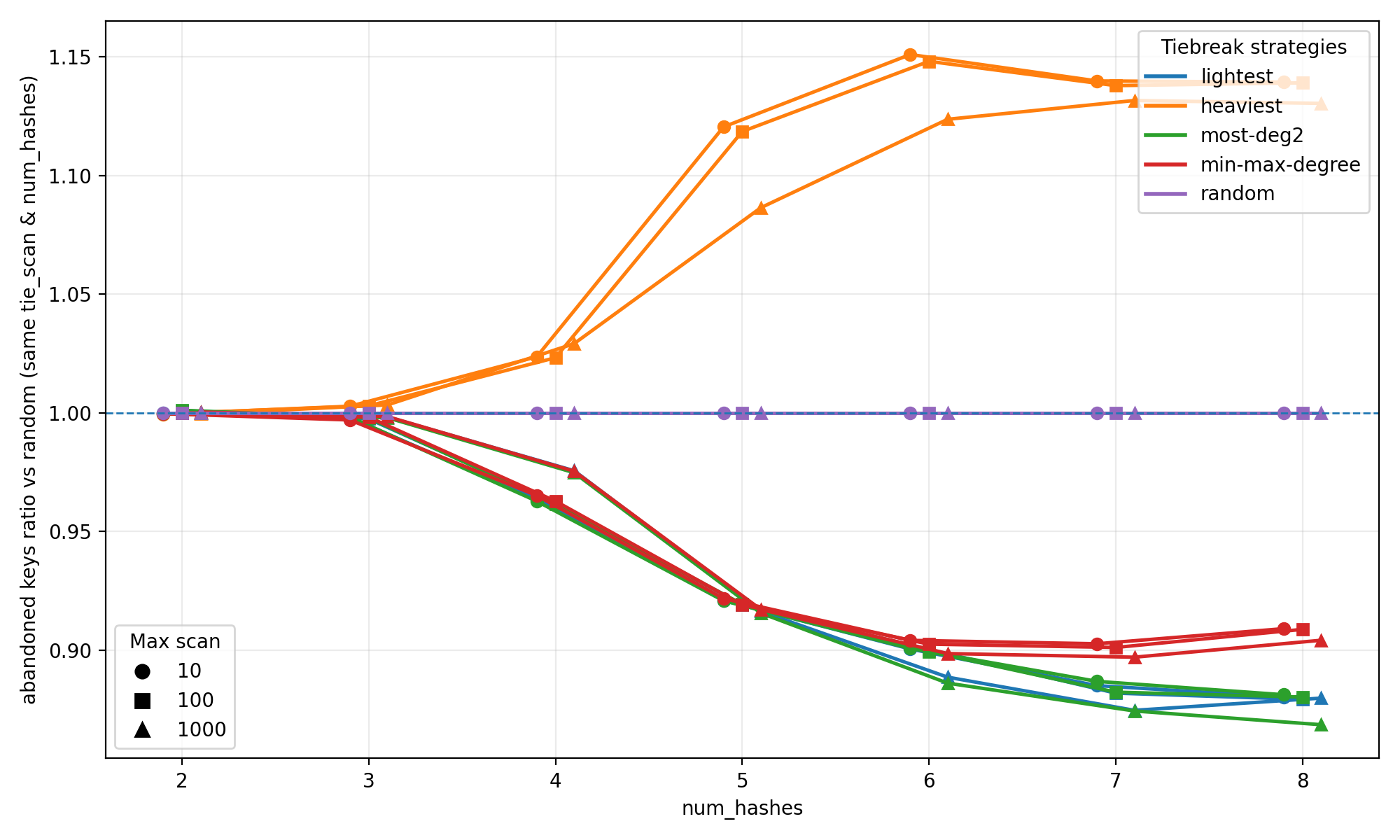}
    \includegraphics[width=0.49\linewidth]{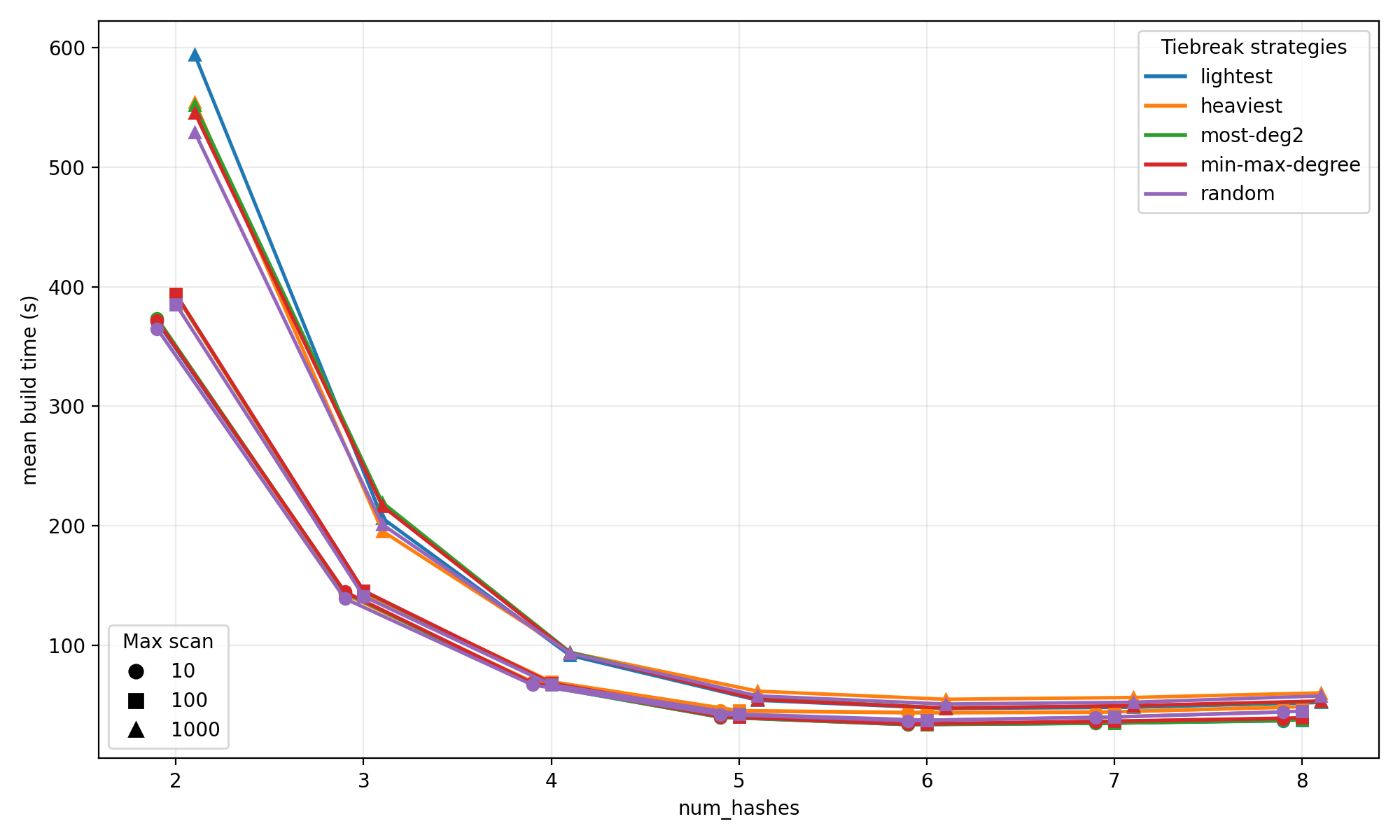}
    \caption{Impact of tie-breaking strategies on abandoned fraction  (left) and construction time (right) for $n=10$M keys.}
    \label{fig:Strategies}
\end{figure}

Next, we evaluate tie-breaking strategies for choosing a cell to treat at blocking events (Figure~\ref{fig:Strategies}). 
Compared to a random choice, the degree-based heuristics reduce the abandoned fraction in our setup, while some strategies can also increase $\alpha$ depending on arity and scan budget. 
Increasing \texttt{T} yields modest additional gains but can noticeably increase build time, especially at low arity.

\begin{figure}
    \centering
    \includegraphics[width=0.49\linewidth]{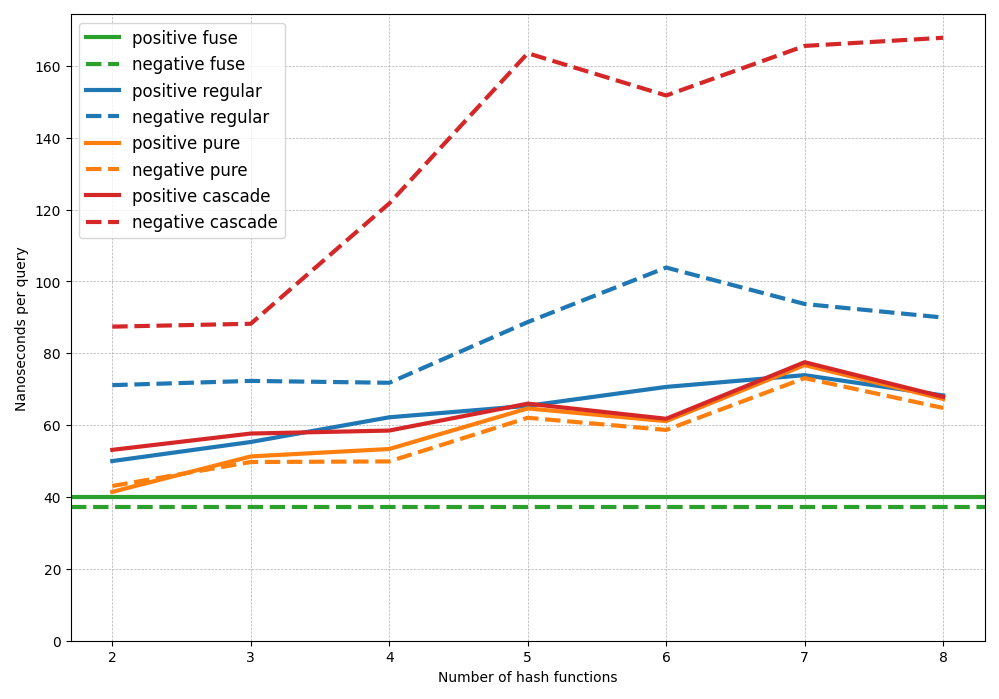}
    \includegraphics[width=0.49\linewidth]{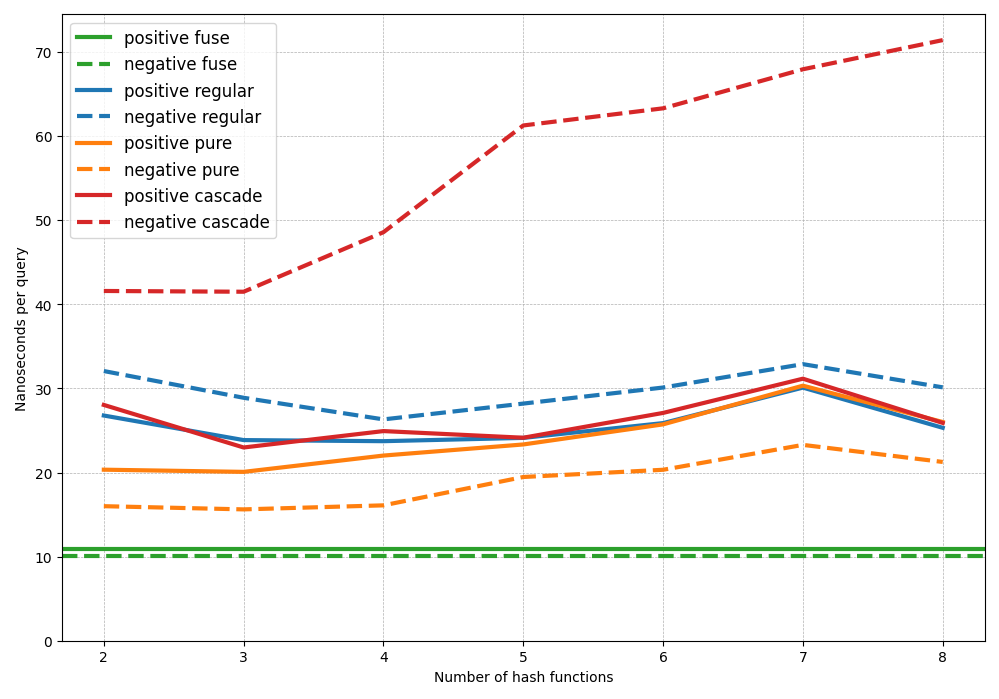}
\caption{Query time for ZOR filters: comparison of positive and negative queries under different arities $N$ without ordering (left) and with ordering (right).}

    \label{fig:query}
\end{figure}

We then evaluate query performance (Figure~\ref{fig:query}, left).
Positive queries are faster than negative queries because negative queries must reject in both the main and auxiliary structures, whereas positive queries often terminate after the first match.
Varying $N$ has a comparatively minor impact on query time in our implementation: after the first random access, subsequent accesses are frequently within nearby segments, and the dominant cost becomes a small number of cache-line touches.
Overall, query times are on the order of $100$\,ns per key in our setup; the additional cost relative to a single-stage fuse filter is concentrated on negative queries due to the auxiliary check.
As expected, cascading auxiliary stages increases query time, particularly for negatives, while pure ZOR behaves similarly to fuse filters (differences are primarily due to low-level optimisations).

To explore a different trade-off, we also evaluate an ordered-query workload in which queries are sorted by segment position so that successive lookups target nearby memory regions, improving cache locality.
This ordering improves absolute query time while largely preserving the relative performance trends across implementations (Figure~\ref{fig:query}, right).

\begin{figure}
    \centering
        \includegraphics[width=0.49\linewidth]{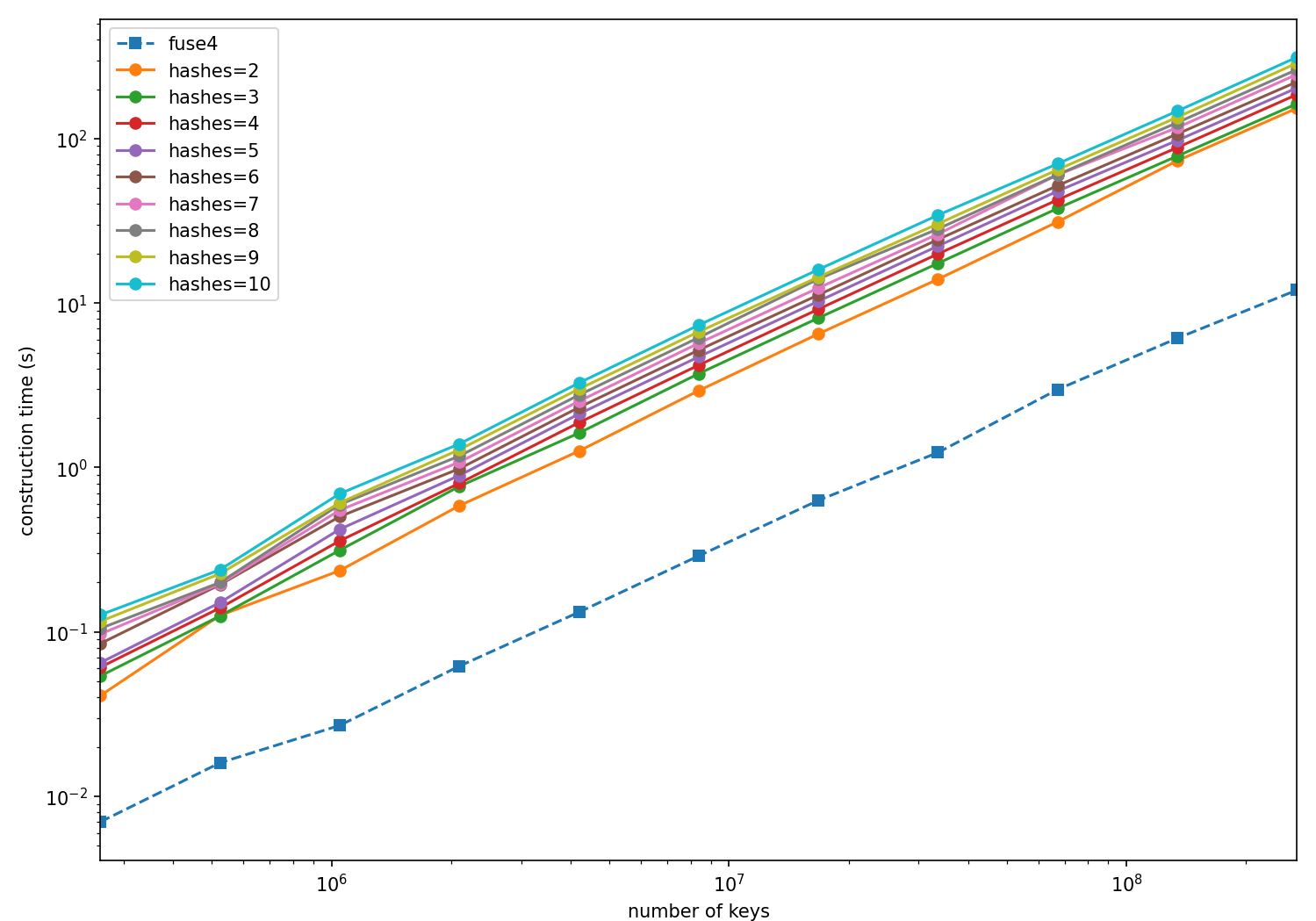}
    \includegraphics[width=0.49\linewidth]{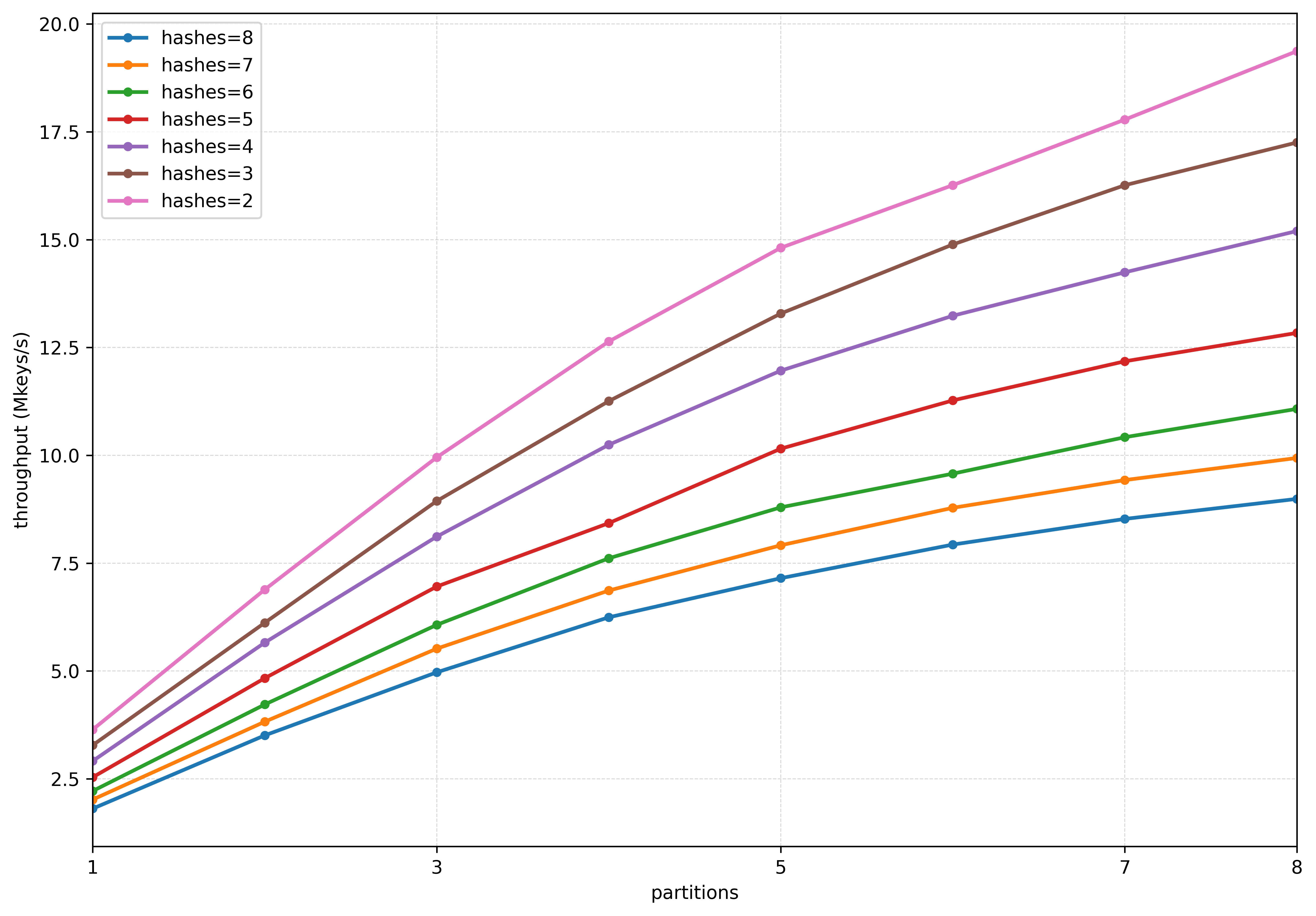}

\caption{Construction time for ZOR filters under varying arity $N$ and set size $n$ (left). Construction time for ZOR filters under varying arity $N$ and partition count $P$ (right).}

    \label{fig:construction}
\end{figure}

Finally, we evaluate construction time (Figure~\ref{fig:construction}, left).
Increasing the arity $N$ increases incidence-processing work and the number of updates during peeling; in our experiments, moving from $N=2$ to $N=8$ roughly doubles construction time. 
Most importantly, our current ZOR prototype is about an order of magnitude slower to build than the highly optimised construction pipeline of Graf and Lemire for fuse filters \cite{graf2022binary}.
Because the core logical stages (segment-local hashing, degree tracking, peeling order, reverse assignment) are closely related, many engineering optimisations should transfer to ZOR.
Closing this gap is the main future implementation goal.
We also evaluate the efficiency of a coarse-grained parallelization strategy that partitions keys into $P$ independent buckets using a hash function and builds the corresponding substructures in parallel (Figure~\ref{fig:construction}, right).
We observe near-linear throughput scaling with the number of partitions/cores in this setting.

\section{Conclusions}
\label{sec:Conclusions}

We presented ZOR filters, a deterministic continuation of XOR and fuse filters that guarantees construction termination while preserving fuse-like query behavior.
More importantly, ZOR achieves near memory optimality at high throughput: in our experiments, we achieve overall overhead below 1\% above the information-theoretic bound while sustaining membership tests on the order of $10^2$\,ns in our setup.

By abandoning a small fraction of keys instead of restarting, ZOR converts probabilistic build failure into an explicit, measurable remainder set $A$ whose size is quantified by the abandoned fraction $\alpha$.
Standard false-positive-only semantics are recovered by storing the remainder in a small auxiliary filter.
A simple space/error balance shows that the auxiliary fingerprint size should scale as $F+\log_2(1/\alpha)$; in practice, a constant offset such as $G=F+8$ is sufficient when $\alpha$ is below the percent scale.
Empirically, abandonment below $1\%$ is routinely achieved for moderate arity, so the auxiliary handles only a tiny fraction of keys and the total space stays extremely close to $\log_2(1/\varepsilon_{\mathrm{tot}})$ bits per key.

The main limitation of the current implementation is construction speed, stemming from the need to maintain explicit incidence information to support deterministic progress when peeling blocks in multi-degree cores.
Future work is therefore primarily engineering-focused: reducing memory traffic during peeling, improving adjacency representations, and adopting segment-local buffering and vectorized hashing in the style of optimised fuse builders.
Algorithmic directions include improved abandonment policies, exploration of sizing choices ($m/n$) as a space/time/abandonment trade-off, query-side optimisations that reduce auxiliary checks for negative queries, and a more systematic study of pure ZOR filters for applications where a tiny false-negative rate is acceptable.

\bibliographystyle{unsrturl}
\bibliography{ref}  

\appendix
\section{Derivation of the memory-optimal auxiliary fingerprint size}
\label{app:aux-opt}

We consider the model from Section~\ref{sec:Methods}.
The main structure stores $F$ bits per original key, and the auxiliary stores $G$ bits per auxiliary-stored key.
If the abandoned fraction is $\alpha$, the total bits per original key are
\[
B(F,G)=F+\alpha G.
\]
Under the uniform fingerprint model, the false-positive probabilities satisfy
\[
\varepsilon_1=2^{-F}, \qquad \varepsilon_2=2^{-G}.
\]
When a query returns ``present'' if either structure matches, the overall false-positive probability is
\[
\varepsilon_{\mathrm{tot}}(F,G)=1-(1-\varepsilon_1)(1-\varepsilon_2)
=\varepsilon_1+\varepsilon_2-\varepsilon_1\varepsilon_2.
\]
When $\varepsilon_1,\varepsilon_2\ll 1$, we use the approximation
\[
\varepsilon_{\mathrm{tot}}(F,G)\approx 2^{-F}+2^{-G}.
\]
Fix a target overall false-positive rate $\varepsilon\in(0,1)$.
We want to minimise $B(F,G)$ subject to the constraint
\[
2^{-F}+2^{-G}=\varepsilon.
\]
Let $a=2^{-F}$ and $b=2^{-G}$, so the constraint is $a+b=\varepsilon$ with $a,b\in(0,\varepsilon)$.
Also, $F=-\log_2 a$ and $G=-\log_2 b$, hence the objective becomes
\[
B = -\log_2 a - \alpha \log_2 b.
\]
Minimizing $B$ is equivalent to minimizing
\[
\widetilde{B} = -\ln a - \alpha \ln b
\]
because $\log_2 z = (\ln z)/(\ln 2)$ differs only by a positive constant factor.

Using the constraint $b=\varepsilon-a$, define
\[
\phi(a)= -\ln a - \alpha \ln(\varepsilon-a), \qquad a\in(0,\varepsilon).
\]
Differentiate:
\[
\phi'(a)= -\frac{1}{a} - \alpha\cdot \frac{-1}{\varepsilon-a}
= -\frac{1}{a} + \frac{\alpha}{\varepsilon-a}.
\]
At an interior optimum, $\phi'(a)=0$, hence
\[
-\frac{1}{a} + \frac{\alpha}{\varepsilon-a}=0
\quad\Longleftrightarrow\quad
\frac{\alpha}{\varepsilon-a}=\frac{1}{a}
\quad\Longleftrightarrow\quad
\alpha a = \varepsilon-a
\quad\Longleftrightarrow\quad
a=\frac{\varepsilon}{1+\alpha}.
\]
Then
\[
b=\varepsilon-a=\varepsilon-\frac{\varepsilon}{1+\alpha}
=\frac{\alpha\varepsilon}{1+\alpha}.
\]
Therefore the optimal split satisfies
\[
b=\alpha a,
\]
i.e.,
\[
2^{-G^\star} \approx \alpha\,2^{-F}.
\]
Taking $-\log_2(\cdot)$ of both sides yields
\[
G^\star \approx F+\log_2\!\Bigl(\frac{1}{\alpha}\Bigr),
\]
which is the stated balance condition.
\qed

\end{document}